\documentclass[final,3p,times,twocolumn]{elsarticle}

\usepackage{graphicx}

\usepackage{amssymb}

\def\lnod{La$_2$NiO$_{4+\delta}$}
\def\lsno{La$_{2-x}$Sr$_x$NiO$_4$}
\def\lsco{La$_{2-x}$Sr$_x$CuO$_4$}
\def\lbco{La$_{2-x}$Ba$_x$CuO$_4$}

\def\lnsco{La$_{1.6-x}$Nd$_{0.4}$Sr$_x$CuO$_4$}

\def\ybco{YBa$_2$Cu$_3$O$_{6+x}$}

\journal{Physica C}

\begin{document}

\begin{frontmatter}



\title{La$_{2-x}$Ba$_x$CuO$_4$ as a superconducting Rosetta Stone}


\author{John M. Tranquada}
\ead{jtran@bnl.gov}

\address{Condensed Matter Physics \&\ Materials Science Division, Brookhaven National Laboratory, Upton, New York 11973-5000, USA}

\begin{abstract}
The high-temperature superconductivity in layered cuprates discovered by Bednorz and M\"uller arrived as a shock.  Gradually, the presence of competing orders, such as antiferromagnetism and charge order, were discovered; however, the relationship to the superconductivity has been confusing.  It so happens that the original cuprate superconductor family \lbco\ contains all of the relevant phases, with extreme competition among them, and analysis of these phases provides strong clues to the nature of the superconductivity in cuprates.
\end{abstract}




\end{frontmatter}


\section{The Early Days}
\label{}

The late K. Alex M\"uller (KAM) effectively determined my career when, with J. Georg Bednorz (JGB), he had the wonderful idea to look for superconductivity in La-Ba-Cu-O \cite{bedn86,bedn88}.  I had just been hired in October 1986 by Myron Strongin as a staff scientist in the Electron Spectroscopy Group of the Physics Department at Brookhaven, as rumors of the discovery, and attempts to replicate it, were gradually filtering through.  Though my assignment was to work on Peter Johnson's inverse photoemission spectrometer, I soon learned that Arnie Moodenbaugh was making samples of superconducting \lsco, and I arranged to do some x-ray spectroscopy measurements on those samples with Steve Heald at the X-11 beam line of the National Synchrotron Light Source (NSLS), the facility that had drawn me to Long Island three years before as a postdoc.

In our first cuprate paper \cite{tran87a}, we found that the Cu$^{2+}$ ions, as probed by x-ray absorption spectroscopy at the Cu $K$ edge, appeared to show no significant valence change on doping, although spectra at the La $L_3$ edge suggested that holes might occur in O $2p$ orbitals (as later confirmed by O $K$-edge studies \cite{chen91}).  This was my first exposure to the strong-correlation effects of the cuprates and I quickly got hooked.  The usual place to present such new results would be the March Meeting of the American Physical Society; however, the avalanche of studies that began to appear came after the deadline for abstract submission for the 1987 meeting, held in New York City.  Fortunately, Arnie learned that Brian Maple was organizing a last-minute session on high-temperature superconductivity, and he got a 5 min.\ slot after midnight in the session that began at 8 pm on a Wednesday evening; he generously gave me half of his time to describe our x-ray results.  What later became known as the ``Woodstock of Physics'' \cite{wood87} was my first chance to be in the same room as KAM, although I had to watch his presentation from the back of a very large ballroom with a couple of thousand other enthusiasts.

\section{Superconductivity from competing interactions}

KAM had long experience with perovskite compounds.  Having a chemical formula of $AB$O$_3$, the $B$ site is typically occupied by a transition-metal ion that is coordinated by six O ions, forming a regular octahedron.  KAM was an expert on compounds such as SrTiO$_3$ \cite{mull79} and was aware of the possibility of superconductivity in perovskites, as had been demonstrated in BaPb$_{1-x}$Bi$_x$O$_3$ with a maximum superconducting transition temperature $T_c=13$~K \cite{slei75}.

For cubic symmetry, the crystal field from the O octahedron is expected to energetically split the five $d$ levels on the $B$ site into two groups, with the higher-energy $e_g$ states including $d_{x^2-y^2}$ and $d_{3z^2-r^2}$.  For ions such as Ni$^{3+}$ ($3d^7$) and Cu$^{2+}$ ($3d^9$), the Jahn-Teller theorem predicts that the octahedra will distort to break the energy degeneracy between the $e_g$ orbitals.

The motivating idea behind the big discovery \cite{bedn88} was to mix such Jahn-Teller ions into a perovskite lattice with the hope that the competition between the local electronically-driven distortions and the average cubic structure would lead to electron pairing and superconductivity.  When studies of LaNiO$_3$ yielded no superconductivity, an attempt was made on a cuprate perovskite, where the superconducting phase that was discovered \cite{bedn86} was found to correspond to the layered perovskite \lbco\  \cite{bedn87}.

The \lbco\ (LBCO) structure contains CuO$_2$ layers in which the Cu sites form a square lattice with O atoms bridging nearest-neighbor Cu sites.  The apical oxygens are distinct, with a considerably longer bond length that automatically satisfies the Jahn-Teller theorem; nevertheless, the idea of competing interactions applies in an unexpected way.  Anderson quickly pointed out that La$_2$CuO$_4$ should be an antiferromagnetic insulator due to strong Coulomb repulsion within the half-filled Cu $3d_{x^2-y^2}$ orbital \cite{ande87}.  The prediction of antiferromagnetism was soon confirmed by neutron diffraction at Brookhaven's High-Flux Beam Reactor (HFBR) \cite{vakn87}.  When holes are doped into the CuO$_2$ planes by substituting Ba$^{2+}$ for La$^{3+}$, their tendency to reduce kinetic energy by delocalizing inevitably competes with the superexchange interaction between neighboring Cu atoms.  The challenge of understanding how that competition can lead to superconductivity has kept many of us busy for nearly four decades.

\section{Turning to neutrons}

As there was no long-term support for my position, Peter Johnson found that Gen Shirane and the Neutron Scattering Group had an opening.  With the discovery of antiferromagnetic order in La$_2$CuO$_4$ \cite{vakn87}, neutron scattering appeared to be a promising adventure, so I joined that group in June 1987.

My appreciation for the significance of the antiferromagnetism developed from interactions with Brookhaven theorist Vic Emery.  Vic had been on sabbatical in Orsay when all of the initial excitement hit; Peter introduced me to him after his return to Brookhaven in early 1987.  Vic had extensive experience with low-dimensional correlated-electron systems, such as organic superconductors.  It was natural to him to consider the CuO$_2$ planes in terms of a Hubbard model with a strong Coulomb repulsion on the Cu sites such that doped holes would tend to go into the O $2p_\sigma$ orbitals.  His paper describing what would become known as the Emery model  was published in the spring of 1987 \cite{emer87}.

After superconductivity was discovered in \ybco\ (YBCO) \cite{wu87}, it seemed likely that a parent compound would be antiferromagnetic (AFM).  In discussing the position in the Neutron group with Shirane, I proposed this idea, and he suggested that it could be my first experiment.  I was able to obtain a sample of YBa$_2$Cu$_3$O$_{6.15}$ from Walter Kunnmann in the Chemistry Department.
I quickly found a possible magnetic peak in powder diffraction, but it did not seem to weaken at 300~K (the N\'eel temperature in the initial La$_2$CuO$_{4+\delta}$ was just 220~K \cite{vakn87}; however, that was lowered from above 300~K due to unappreciated interstitial oxygen \cite{yama87,well97}), and it was not trivially indexed.  It was only after I was able to get a partially oriented powder (grains with $c$ axis aligned in a strong magnetic field due to anisotropy of the magnetic susceptibility) prepared by Mas Suenaga's group that I was able to identify that peak as $(\frac12,\frac12,1)$.  As the $(\frac12,\frac12,0)$ was absent, it became clear that the AFM order in neighboring planes of the bilayers had to be antiparallel; with that realization, I quickly wrote up the results.  With only one magnetic peak, I could not determine the spin direction, so I naively assumed that it was along the $c$ axis.  After submitting the paper to PRL, I got some time on the triple-axis spectrometer on the cold neutron source, where I was able to resolve the $(\frac12,\frac12,2)$ peak from the (0,0,3), and its intensity indicated that the spins had to be in plane.  I was able to get that correction into the paper before it was published \cite{tran88a}.

\section{Structures}

My connection to KAM was an indirect one through Vic Emery and John Axe.  Emery had previously been a Visiting Scientist at the IBM Z\"urich Research Laboratory.  Axe had worked at the IBM T. J. Watson Research Center in the 1960's, before joining the Neutron Scattering Group at Brookhaven in 1970.  His first neutron scattering paper was a study of soft phonons in LaAlO$_3$ with KAM and Gen Shirane \cite{axe69}, with a follow up study by the latter two a few years later \cite{kjem73}.  There were also shared interests in SrTiO$_3$ \cite{shir69,shap72,mull79}.  The neutron scattering studies focused on structural transitions involving rotations of the oxygen octahedra and associated soft phonons.

\begin{figure}[t]
 \centering
    \includegraphics[width=\columnwidth]{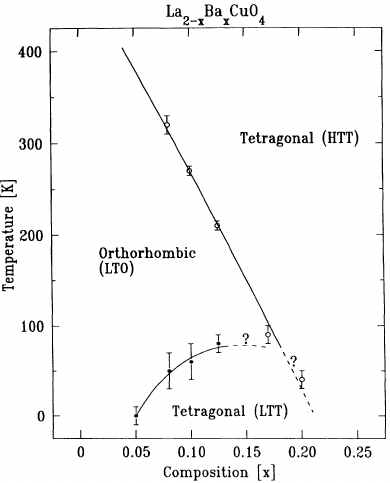}
    \caption{\label{fg:struc} Phase diagram for crystal structure of LBCO determined by x-ray diffraction.  Reprinted with permission from Ref.~\cite{axe89}, \copyright 1989 by the American Physical Society. }
\end{figure}

In the summer of 1988, KAM and JGB organized a workshop in Oberlech, Austria through the IBM Europe Institute.  Emery and Axe were invited speakers and I was able to tag along.  Axe reported on x-ray and neutron diffraction studies that revealed a series of structural phase transitions in LBCO involving tilt displacements of the CuO$_6$ octahedra \cite{axe89b}. After repeating such studies at the NSLS for a range of compositions, he published the structural phase diagram shown in Fig.~\ref{fg:struc}.  The structure starts in the high-temperature tetragonal (HTT) phase of K$_2$NiF$_4$.  On cooling, the octahedra displace in a staggered tilting pattern about a diagonal axis with respect to the in-plane Cu-O bonds, resulting in the low-temperature orthorhombic structure (LTO) that is also exhibited by \lsco\ (LSCO).  The surprise was a further transition to a low-temperature tetragonal (LTT) phase, in which the octahedra in one layer rotate about one set of Cu-O bonds, and the pattern is rotated by $90^\circ$ in the neighboring layers.

\begin{figure}[t]
 \centering
    \includegraphics[width=\columnwidth]{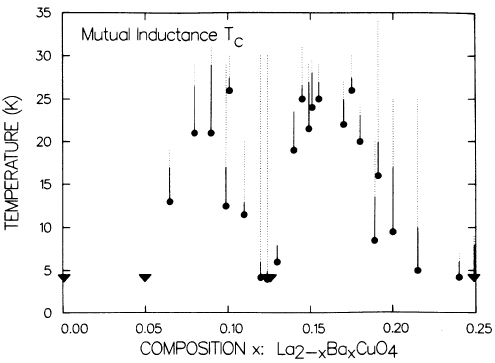}
    \caption{\label{fg:tc} Superconducting transition temperature vs.\ Ba content in \lbco\ measured by Moodenbaugh and coworkers on polycrystalline samples.  Reprinted with permission from  \cite{mood88}, \copyright 1988 by the American Physical Society. }
\end{figure}

In the meantime, Moodenbaugh and coworkers \cite{mood88} had discovered a surprising dip in $T_c$ in LBCO at $x\sim\frac18$, as shown in Fig.~\ref{fg:tc}.  Much of the superconductivity and the anomalous dip occur within the LTT phase.  Could there be a connection?

To make progress with neutron scattering, large single crystals were needed.  LBCO was particularly challenging to grow in single crystal form at interesting dopant concentrations.
The first cuprate crystals were obtained for LSCO.  Neutron scattering studies showed that hole doping led to the rapid elimination of AFM order, but some sort of incommensurate AF correlations were found to survive at low excitation energies \cite{kast98,yama98a}. 

KAM and JGB organized another workshop in the summer of 1989, this time in Erice, Sicily.  At this meeting, I was able to contribute on what we had learned about AFM order and the impact of CuO chain ordering in YBCO \cite{tran90}.  Again, it was a great chance for an early-career researcher to meet and interact with leading figures in the high-$T_c$ field.

\section{To nickelates and back to cuprates}

The HFBR began routine operations at a power of 40 MW in 1966.  A decision was made to raise the power to 60 MW in 1982, 5 years before I began to use it for experiments.  The HFBR was temporarily shut down in 1989 to reconsider the effectiveness of cooling of the core in the case of an emergency, eventually being restarted at 30 MW in 1991.  It then operated through the end of 1996, but never again after that \cite{crea22}.  That final operating period became a crucial time of discovery for me.

Around 1992, Doug Buttrey (University of Delaware) proposed to study the behavior of \lnod\ and \lsno, compounds that are isostructural to cuprates and for which he had grown sizable single crystals.  While it was known that the parent material was an antiferromagnetic insulator \cite{land89}, studies of polycrystalline \lsno\ demonstrated that it was possible to induce a metallic phase by doping, but there was no sign of superconductivity \cite{cava91}.  

\begin{figure}[t]
 \centering
    \includegraphics[width=\columnwidth]{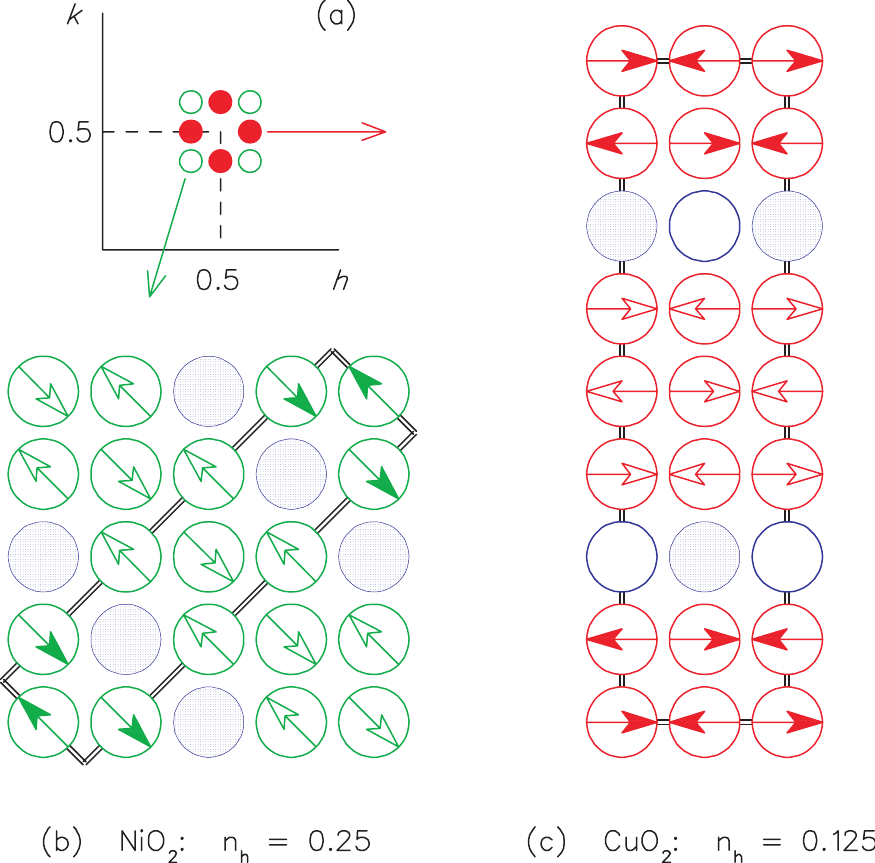}
    \caption{\label{fg:diag} Comparison of (a) magnetic superlattice peaks associated with (b) diagonal stripe order in \lsno\ with $x=0.25$ and (c) bond-parallel stripe order in \lnsco\ with $x=0.125$.  Circles indicate Ni/Cu sites; arrows indicte orientation of ordered magnetic moments; shaded circles indicate presence of a single doped hole.  Adapted from \cite{tran95a}. }
\end{figure}

Around this time, Jan Zaanen gave a seminar at Brookhaven on a theoretical analysis suggesting that doped holes in LSCO would localize as polarons \cite{anis92}.  I had this in mind as we started our experiments on \lnod.
We began by studying the ordering of oxygen interstitials and the phase separation into distinct phases of interstitial density \cite{tran94b}.  Then we discovered a higher-density phase that seemed to have three-dimensional (3D) interstitial order and in which the antiferromagnetically-correlated Ni spins and the doped holes ordered as intertwined stripes \cite{tran94a}, as indicated schematically in Fig.~\ref{fg:diag}(a).  Other groups had recently found evidence for the incommensurate magnetic scattering \cite{hayd92} and charge order \cite{chen93} in distinct samples, but our measurements were the first to demonstrate that these modulations were mutually commensurate, as supported by subsequent calculations \cite{zaan94}.

The idea of spin and charge stripe correlations in cuprates was originally discovered in Hartree-Fock calculations for CuO$_2$ planes \cite{zaan89,mach89}; however, these insulating stripes had a period that was inconsistent with the incommensurability of AFM spin excitations measured in LSCO crystals, and so were dismissed \cite{cheo91}.  Independently, Vic Emery was working with Steve Kivelson on the concept of frustrated electronic phase separation \cite{emer93}, which could lead to modulated structures that included stripes with a range of periods \cite{low94}.

The dip in $T_c$ for LBCO at $x=1/8$ might be due to stripe order, but a single-crystal sample was not available to test this possibility.  Fortunately, there was an alternative: La$_{1.6-x}$Nd$_{0.4}$Sr$_x$CuO$_4$ (LNSCO) also shows a dip in $T_c$ at $x=0.12$ and it shows the same structural phases as LBCO \cite{craw91}; furthermore, Uchida's group at the University of Tokyo had grown crystals of this system \cite{naka92}.

Axe got a small crystal of LNSCO $x=0.12$ from Uchida as an analog to LBCO to study the structural transitions by neutron diffraction.  Based on our La$_2$NiO$_{4+\delta}$ results, I had an idea of what to look for in terms of possible stripe order.  In a collaborative neutron diffraction experiment, I quickly found the incommensurate AF elastic peaks, and then had to count overnight to see the charge order peaks.  The corresponding stripe order is indicated in Fig.~\ref{fg:diag}(c); the hole density in the charge stripes is one hole for every two Cu sites along the stripe, consistent with metallic behavior. I proceeded to write up the results and submitted to Nature \cite{tran95a}.
Fujita and Yamada eventually had success in growing crystals of LBCO at $x=1/8$, which allowed confirmation of the charge and spin stripe order by neutron scattering \cite{fuji04}.  After we hired Genda Gu at Brookhaven, he grew a sequence of LBCO crystals, enabling Markus H\"ucker to investigate much of the phase diagram \cite{huck11}.

\section{Coupling to structure}

The dependence of $T_c$ on material properties can provide important clues to the nature of the electron pairing.  For example, an effective attractive interaction in conventional superconductors is provided by phonons \cite{bard57}.  An early experimental observation consistent with $T_c \sim 1/M^{0.5}$ obtained by using different isotopes to vary the atomic mass $M$ in an elemental superconductor \cite{alle50} provided support for such a mechanism, as the phonon frequencies are proportional to $1/\sqrt{M}$.

Given that his original motivation was based on the concept of polarons, KAM had a strong interest in the isotope effect and wrote a review of early results, where the mass exponent for oxygen isotopes was much less than a half \cite{mull90}; he also participated in a later review \cite{bish07} that emphasized the care that must be taken in determining sample stoichiometry in order to obtain a proper measure of the isotope effect.  It was just while he was preparing the first review that exciting results started to appear from Mike Crawford and coworkers on LSCO and LBCO with an exponent $>0.5$ in the underdoped regime \cite{craw90a,craw90b}.  Later work on LSCO at more concentrations by KAM and coworkers \cite{zhao98} demonstrated that the maximum exponent occurs near $x\sim0.12$, where we now know that stripe order occurs \cite{crof14}. 

More recent measurements on LBCO $x=1/8$ indicated that the strong effect is a consequence of the sensitivity to the oxygen mass of the structural transition from LTO to LTT and the associated stripe ordering \cite{gugu14,gugu15}.  Combining this sensitivity with the competition between bulk superconductivity and stripe order \cite{tran97a} leads to the anomalously large isotope effect.

KAM had recognized the role of inhomogeneity on the development of superconducting order in early LBCO samples \cite{mull87}, so it was not a stretch for KAM and collaborators to appreciate the role that stripe correlations could play \cite{buss01,miha02,shen15}.  The emphasis tended to be on coupling to phonons and the lattice, especially the formation of bipolarons \cite{kell08}.  While there is direct evidence of phonon anomalies consistent with screening of charge modulations \cite{rezn06,peng20,wang21a}, as well as static lattice modulations that screen charge stripes \cite{sear23}, the interaction of dopant-induced holes with antiferromagnetically-correlated Cu moments has dramatic effects \cite{fuji12a} that provide a possible explanation for the superconductivity \cite{tran21a}.

\section{Stripe order and 2D superconductivity}

While stripe order clearly competes with bulk, 3D superconductivity \cite{tran97a,huck11}, measurements by Qiang Li and Markus H\"ucker on LBCO $x=1/8$ provided evidence of 2D superconductivity, with a mean-field transition at $\sim40$~K, together with the spin-stripe ordering, and a phase ordering (Berezinskii-Kosterlitz-Thouless) transition at 16~K \cite{li07,tran08}.  One surprise here is that 2D superconducting order can develop without an interlayer Josephson coupling leading to 3D order.  To explain the apparent frustration of the interlayer Josephson coupling, the concept of pair-density-wave (PDW) superconductivity was proposed \cite{berg07,agte20}.  The idea here is that the pair wave function is positive on one charge stripe and negative on the next.  Because the charge stripe orientation rotates $90^\circ$ from one layer to the next, following the LTT lattice symmetry, the interlayer Josephson coupling averages to zero. 

Spectroscopic characterizations of LBCO $x=1/8$ have not identified any features that would suggest that the pairing mechanism in stripe-ordered cuprates is distinct from that in samples with spatially-uniform superconductivity.  Hence, it appears that intertwined orders may be a common feature of hole-doped cuprates \cite{frad15}.  It should be noted, of course, that the concept of pairing within stripes does not necessarily require static stripe order; dynamic stripe correlations that are defined on an energy scale smaller than the local pairing energy may still yield uniform superconductivity.

\section{Pairing in charge stripes}

\begin{figure}[t]
 \centering
    \includegraphics[width=\columnwidth]{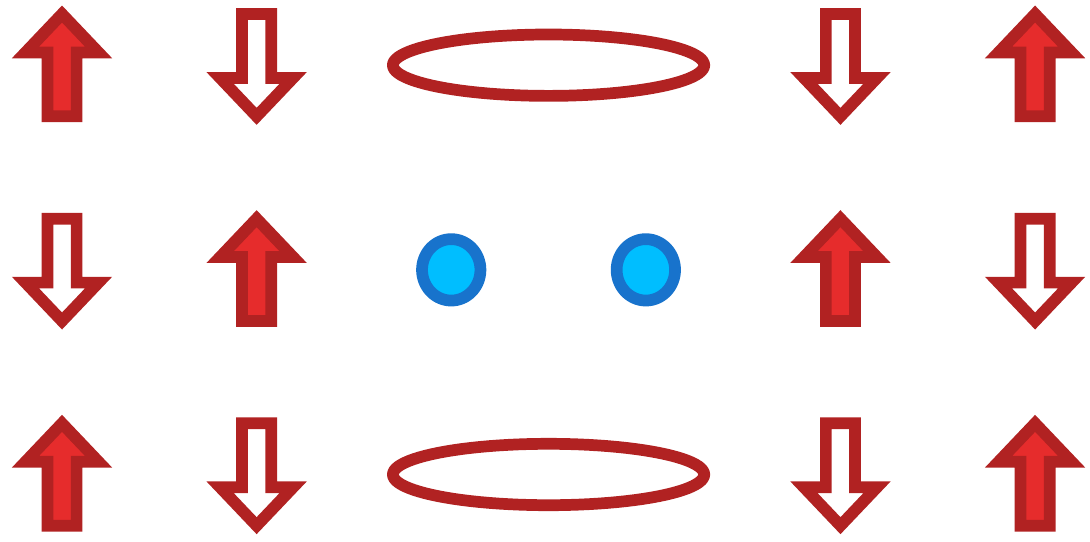}
    \caption{\label{fg:SC} Schematic of proposed pairing concept in LBCO.  Arrows indicate ordered Cu moments in spin stripes; ellipses indicate spin-singlet correlations between pairs of sites within the charge stripe; blue circles are holes that pair in order to avoid causing singlet-to-triplet excitations within the charge stripe.  From \cite{tran21a}. }
\end{figure}

The observation of 2D superconductivity in stripe-ordered LBCO $x=1/8$ suggests that pairing can occur within the charge stripes.  Further support for this idea comes from measurements of the same material in high magnetic fields (applied along the $c$ axis), where 35 T results in an unusual metallic phase with a resistivity twice the quantum of resistance for pairs and a Hall constant that is negligible (below 16 K) suggesting particle-hole symmetry survives even in the absence of superconductivity \cite{li19a}.

The antiphase spin stripes that define the charge stripes also create a geometric frustration for the spin degrees of freedom within the charge stripes.  From such a perspective \cite{tran21a}, it is natural to think of the charge stripes as hole-doped two-leg spin ladders.  An undoped spin-$\frac12$ ladder has a singlet-triplet excitation gap as large as $J/2$, where $J$ is the superexchange energy \cite{barn93,dago96}.  For $J\approx 100$~meV, the gap can be as large as 50 meV, the size of the gap observed at the AFM wave vector in LBCO $x=1/8$ \cite{tran04}.  When holes are doped into such a ladder, they will pair up in order to avoid breaking the spin singlet correlations \cite{dago92,dago96}, as illustrated in Fig.~\ref{fg:SC}.  In order to get 2D superconducting order, it is necessary to establish phase order between neighboring charge stripes \cite{emer97}.  The spin stripes get in the way of uniform phase order, but with a $\pi$ phase shift, one obtains PDW order.

To get uniform superconducting order, it is necessary to gap the spin stripes.  Such a gapping of the spin-stripe excitations on cooling below $T_c$ is commonly observed by neutron scattering, with the LSCO system being a good example \cite{lake99}.  Comparisons with the literature reveal that the magnitude of the spin gap in LSCO is equal to the coherent SC gap and that the spin gap, in general, provides an upper limit to the coherent SC gap \cite{li18}.  As the spin correlations weaken with overdoping \cite{waki07b}, the pairing strength and superconducting order also disappear \cite{li22}.

While the picture discussed here is a speculative interpretation of empirical results, some support for it comes from recent numerical analyses of the Hubbard model \cite{jian21,mai22,pons23,maie20,jian23}.

\section{Conclusion}

KAM had a brilliant idea to test whether competing interactions might lead to high-temperature superconductivity.  Moreover, the very LBCO compound that he originally discovered contains anomalous features that, much like a Rosetta Stone, allow us to identify and translate the key features for pairing relevant to cuprate families with much higher $T_c$'s.  I am grateful to Alex M\"uller for discovering a puzzle that has kept me happily occupied for most of life, as well as to mentors such as Vic Emery and John Axe, who provided invaluable guidance and connections.

\section{Acknowledgment}

Work at Brookhaven is supported by the Office of Basic Energy Sciences, Materials Sciences and Engineering Division, U.S. Department of Energy under Contract No.\ DE-SC0012704. 





\bibliographystyle{elsarticle-num}
\bibliography{lno,theory}







\end{document}